\documentclass[aps,prd,preprint,showpacs]{revtex4}

\usepackage{longtable}
\usepackage{textcomp}
\usepackage{multirow}
\begin{document}
\title{Pion- and strangeness-baryon $\sigma$ terms in the extended chiral
constituent quark model}

\author{C. S. An$^{1}$}\email{ancs@swu.edu.cn}

\author{B. Saghai$^{2}$}\email{bijan.saghai@cea.fr}

\affiliation{1. School of Physical Science and Technology, Southwest University,
              Chongqing 400715, People's Republic of China\\
2. Universit\'e Paris-Saclay, Institut de Recherche sur les lois Fondamentales de l'Univers,
Irfu/SPhN, CEA/Saclay, F-91191 Gif-sur-Yvette, France}

\thispagestyle{empty}

\date{\today}

\begin{abstract}
%
%
Within an extended chiral constituent quark formalism, we investigate
contributions from all possible five-quark components in the octet baryons  to the
pion-baryon ($\sigma_{\pi B}$) and
strangeness-baryon ($\sigma_{s B}$) sigma terms; $B \equiv N,~\Lambda,~\Sigma,~\Xi$.
The probabilities of the quark-antiquark components in the ground-state baryon octet wave functions
are calculated by taking the baryons to be admixtures of three- and five-quark components, with the
relevant transitions handled {\it via} the $^{3}$P$_{0}$ mechanism.
Predictions for $\sigma_{\pi B}$ and $\sigma_{s B}$ obtained by using input parameters taken from 
the literature are reported.
Our results turn out to be, in general, consistent with the findings via
lattice QCD and chiral perturbation theory.

\end{abstract}


\maketitle

%
\section {Introduction}
The pion-baryon sigma terms ($\sigma_{\pi B}$) provide critical information on the nature of
explicit chiral symmetry breaking in QCD, the extent of the breaking, and the decomposition
of the mass of the ground-state baryons, which can be related to the quark-antiquark ($Q \bar{Q}$)
components in the baryons, with $Q \bar{Q} \equiv u \bar{u},~d \bar{d},~s \bar{s}$.
Also of great interest are the strangeness-nucleon sigma terms ($\sigma _{s N}$) arising exclusively
from the strange sea quark-antiquark pairs.

The pion-nucleon $\sigma$ term is the most studied one since the late
1960s~\cite{Reya:1974gk,Gasser:2000wv} and its seminal value was reported in
Ref.~\cite{Gasser:1990ce}: $\sigma_{\pi N}$=45$\pm$8 MeV,
which was obtained by analysis of the $\pi N$ data available in the late 1980s and by taking
into account the current algebra result generated by the quark masses.
As discussed later, various results reported in the literature agree with that canonical value
within 2$\sigma$.
A recent comprehensive study~\cite{Chen:2012nx}, within the covariant
baryon chiral perturbation theory and using SAID phase shift analysis~\cite{Arndt:2006bf}
as well as the lattice quantum chromodynamics (LQCD) data published in 2009 and 2011 led
to $\sigma_{\pi N}$=45$\pm$6 MeV.
However, the extraction of $\sigma_{\pi N}$ from the $\pi N$ scattering data appears
still to be challenging~\cite{Hite:2005tg,Stahov:2012ca,Matsinos:2013jda,Hoferichter:2012wf};
for a comprehensive recent account on the $\pi N$ phenomenology, see Ref.\cite{Alarcon:2012kn}.

With respect to the strangeness sigma terms $\sigma_{s B}$, there is no experimental guidance.
However, a recent work~\cite{Gubler:2014pta}, based on a QCD sum rule approach, put forward a
(almost) linear correlation between the $\phi$-meson mass-peak shift in nuclear matter and the
value of $\sigma _{s N}$.

On the other hand, for more than a decade intensive theoretical efforts have been deployed
to predict the sigma terms.
Those investigations are based on chiral perturbation theory, chiral extrapolation,
and the chiral Lagrangian, often using LQCD data: see, e.g.,
Refs.~\cite{Semke:2012gs,Lutz:2014oxa,Ren:2013dzt,Ren:2014vea,Alarcon:2012nr,Young:2009zb,Shanahan:2012wh,MartinCamalich:2011py,Alvarez-Ruso:2013fza}
and references therein.
A general trend in those results is reaching values for $\sigma_{\pi N}$ around 55 MeV.

In parallel, several collaborations are refining the LQCD calculations and putting forward predictions
for sigma terms; see, e.g., Refs.~\cite{Bali:2011ks,Durr:2011mp,Horsley:2011wr,Alexandrou:2013uwa}
and references therein.
The LQCD results for $\sigma_{\pi N}$ turn out to be mostly in the range of 30-40 MeV.

In the present paper we focus on the quark-antiquark $Q \bar{Q}$ components of the intrinsic
five-quark Fock states in the baryon wave functions, at the origin of nonperturbative phenomena;
for a recent review see Ref.~\cite{Chang:2014jba}.
Our formalism is based on an extended chiral constituent quark approach and embodies
all possible five-quark mixtures in the baryon wave functions, with the mechanism of transition
between three- and five-quark components in the baryons treated within the $^{3}$P$_{0}$
quark-antiquark creation frame~\cite{Le Yaouanc:1972ae,Le Yaouanc:1973xz,Kokoski:1985is}.

Actually, in a series of papers~\cite{An:2010wb,An:2011sb,An:2012kj,An:2013daa} we successfully studied
contributions from the genuine five-quark components in the baryons to various observables,
namely, the hadronic and electromagnetic decays of baryon resonances~\cite{An:2010wb,An:2011sb},
flavor asymmetry of the nucleon sea $\bar d-\bar u$~\cite{An:2012kj} and the strangeness magnetic form
factor of the proton~\cite{An:2013daa}.

In this paper we extend our studies to the pion-baryon sigma terms ($\sigma_{\pi B}$) and
strangeness-baryon ones ($\sigma_{s B}$); $B \equiv N,~\Lambda,~\Sigma,~\Xi$.

The present manuscript is organized in the following way: in Sec.~\ref{sec:Theo}, after a
brief presentation of the theoretical frame, we give explicit expressions for sigma terms relating
them to the quark-antiquark pair probabilities.
Numerical results for $\sigma_{\pi B}$ and $\sigma_{s B}$ are reported in Sec.~\ref{sec:Res} and
compared to the outcomes from other approaches referred to above.
Finally, Sec.~\ref{sec:conclu} contains a summary and conclusions.
Quark-antiquark probabilities in the five-quark components for all
four ground-state octet baryons are tabulated in Appendix A.

%
\section{Theoretical frame}
\label{sec:Theo}
The content of our extended chiral constituent quark model (E$\chi$CQM) was developed
in Refs.~\cite{An:2012kj,An:2013daa}. Hence, in Sec.~\ref{sec:recall}, we briefly report on the
main features of the formalism. In Sec.~\ref{sec:proba} we give explicit expressions for the light
and strange quark-antiquark pairs probabilities in the octet baryons.
Sigma terms are then expressed in terms of those probabilities in Sec.~\ref{sec:s-t}.
\subsection{Chiral constituent quark approach}
\label{sec:recall}
The wave function for the baryon $B$ is expressed as
\begin{equation}
 |\psi\rangle_{B}=\frac{1}{\mathcal{\sqrt{N}}}{\Big[}|qqq\rangle+
 \sum_{i,n_{r},l}C_{in_{r}l}|qqq(Q \bar{Q}),i,n_{r},l\rangle {\Big ]}\,,
\label{wfn}
\end{equation}
where the first term is the conventional wave function for the baryon with three
constituent quarks ($q \equiv u,~d$) and the second term is a sum over all possible higher Fock
components with a $Q \bar{Q}$ pair; $Q \bar{Q} \equiv u \bar{u},~d \bar{d},~s \bar{s}$.
Different possible orbital-flavor-spin-color configurations of the four-quark
subsystems in the five-quark system are numbered by $i$; $n_{r}$ and $l$ denote
the inner radial and orbital quantum numbers, respectively,
while $C_{in_{r}l}/\sqrt{\mathcal{N}}$ represents the
probability amplitude for the corresponding five-quark component.

The coefficient $C_{in_{r}l}$ for a given five-quark component can be related
to the transition matrix element between the  three- and five-quark configurations
of the studied baryon. To calculate the corresponding transition matrix element,
we use a $^{3}$P$_{0}$ version for the transition coupling operator $\hat{T}$,
\begin{eqnarray}
 \hat{T}&=&-\gamma\sum_{j}\mathcal{F}_{j,5}^{00}\mathcal{C}_{j,5}^{00}C_{OFSC}
 \sum_{m} \langle1,m;1,-m|00\rangle\chi^{1,m}_{j,5}\nonumber \\
&&\mathcal{Y}^{1,-m}_{j,5} (\vec{p}_{j}-\vec{p}_{5})b^{\dag}(\vec{p}_{j})d^{\dag}(\vec{p}_{5})\,,
\label{op}
\end{eqnarray}
with $\gamma$ a dimensionless constant of the model,
$\mathcal{F}_{i,5}^{00}$ and $\mathcal{C}_{i,5}^{00}$
the flavor and color singlet of the quark-antiquark pair $Q_{i} \bar{Q}$
in the five-quark system, and $C_{OFSC}$ an operator to calculate the
orbital-flavor-spin-color overlap between the residual three-quark configuration
in the five-quark system and the valence three-quark system.

The probability of the sea quark-antiquark pairs in the baryon $B$ and the normalization
factor read, respectively,
\begin{eqnarray}
\mathcal{P}_B^{Q \bar{Q}}&=& \frac{1}{\mathcal{N}}
\sum_{i=1}^{17}\Bigg[ \Big ( \frac{T_i^{Q \bar{Q}} }{M_B-E_i^{Q \bar{Q}}} \Big )^2 \Bigg ], \\ [10pt]
\label{prob}
\mathcal{N}  &\equiv&  1+ \sum_{i=1}^{17} \mathcal{N}_i
=1+\sum_{i=1}^{17} \sum_{Q \bar{Q}} \Bigg[ \Big ( \frac{T_i^{Q \bar{Q}}}{M_B-E_i^{Q \bar{Q}}}\Big )^2  \Bigg ].
\label{norm}
\end{eqnarray}
where the first term in Eq.~(\ref{norm})~is due to the valence three-quark state,
while the second term comes from the five-quark mixtures.

The probabilities $\mathcal{P}_B^{ Q \bar{Q}}$ provide us with all needed matrix elements to
extract the $\sigma$ terms.

Finally, the 17 possible different five-quark configurations can be classified in
four categories (Table~\ref{caco}) with respect to the orbital and spin wave functions of the
four-quark subsystem, with characteristics summarized below, using the shorthand notation for
the Young tableaux.

{ $\bf i)~[31]_{X}$ {\bf and} $\bf [22]_{S}$:}
The total spin of the four-quark subsystem is $0$.

{$\bf ii)~[31]_{X}$ {\bf and} $\bf [31]_{S}$:}
The total spin of the four-quark subsystem is $1$, combined to the orbital angular momentum
$L_{[31]_{X}}=1$, the total angular momentum of the four-quark subsystem can be $J=0,1,2$,
and to form the proton spin $1/2$, only the former two are possible alternatives.
In the present case, we take the lowest one $J=0$.

{$\bf iii)~[4]_{X}$ {\bf and} $\bf [22]_{S}$:}
 That the total angular momentum of the four-quark subsystem is $0$.

{$\bf iv)~[4]_{X}$ {\bf and} $\bf [31]_{S}$:}
The total spin of the four-quark subsystem should be $S_{[31]}=1$; here we assume that the
combination of $S_{[31]}$ with the orbital angular momentum of the antiquark leads to
$J=S_{4}\oplus L_{\bar{q}}=0$.
%
\begin{table}[hbt]
\caption{\footnotesize Categories (2$^{nd}$ line) and configurations (lines 3-8) for five-quark
components.
\label{caco}}
%
\begin{tabular}{ccccccccccc}
\hline\hline
i & Category / Config. && i & Category / Config. && i & Category / Config. && i & Category / Config.  \\
  & I) $[31]_{X}[22]_{S}$: &&  & II) $[31]_{X}[31]_{S}$ &&  & III) $[4]_{X}[22]_{S}$ &&  & IV) $[4]_{X}[31]_{S}$  \\
\hline
 1 & $[31]_{X}[4]_{FS}[22]_{F}[22]_{S}$    &&  5 & $[31]_{X}[4]_{FS}[31]^1_{F}[31]_{S}$ &&
 11 & $[4]_{X}[31]_{FS}[211]_{F}[22]_{S}$ &&  14 &$[4]_{X}[31]_{FS}[211]_{F}[31]_{S}$ \\
 2 & $[31]_{X}[31]_{FS}[211]_{F}[22]_{S}$  && 6 & $[31]_{X}[4]_{FS}[31]^2_{F}[31]_{S}$ &&
 12 & $[4]_{X}[31]_{FS}[31]^1_{F}[22]_{S}$&&  15 & $[4]_{X}[31]_{FS}[22]_{F}[31]_{S}$ \\
 3 & $[31]_{X}[31]_{FS}[31]^1_{F}[22]_{S}$ && 7 & $[31]_{X}[31]_{FS}[211]_{F}[31]_{S}$ &&
 13 & $[4]_{X}[31]_{FS}[31]^2_{F}[22]_{S}$ &&  16 & $[4]_{X}[31]_{FS}[31]^1_{F}[31]_{S}$ \\
 4 & $[31]_{X}[31]_{FS}[31]^2_{F}[22]_{S}$ && 8 & $[31]_{X}[31]_{FS}[22]_{F}[31]_{S}$ && & &&  17 & $[4]_{X}[31]_{FS}[31]^2_{F}[31]_{S}$ \\
   &                                       && 9 & $[31]_{X}[31]_{FS}[31]^1_{F}[31]_{S}$ && & && & \\
   &                                       && 10 & $[31]_{X}[31]_{FS}[31]^2_{F}[31]_{S}$ && & && & \\
\hline
\hline
\end{tabular}
\end{table}
%
%
\subsection{Quark-antiquark pair probabilities}
\label{sec:proba}
Starting from Eq.~(\ref{prob}), we give explicit expressions for the light quark-antiquark pairs
($u \bar{u}$ and $d \bar{d}$) probabilities for the octet baryon components ($\mathcal{P}_B^{q \bar{q}})$
in terms of the five-quark probabilities per configuration ($P_B(i)$, $i$=1-17) and the relevant squared
Clebsch-Gordan coefficients ($[CG]_{jk}$), i.e.
\begin{equation}
[CG]_{11} = 1,~[CG]_{12} = 1/2,~[CG]_{13} = 1/3,~[CG]_{14} = 1/4,~[CG]_{23} = 2/3,~[CG]_{34} = 3/4.
\end{equation}

In the present work, the probability amplitudes are calculated within the most commonly
accepted $Q\bar{Q}$ pair creation mechanism, namely, the $^{3}$P$_{0}$ model.
Then, the $Q\bar{Q}$ pair is created anywhere in space with the quantum numbers of the
QCD vacuum $0^{++}$, corresponding to $^{3}$P$_{0}$~\cite{Le Yaouanc:1972ae}.
This model has been successfully applied to the decay of mesons and baryons
~\cite{Le Yaouanc:1973xz,Kokoski:1985is}, and has recently been employed to analyze the
sea flavor content of the ground states of the $SU(3)$ octet baryons~\cite{An:2012kj} by
taking into account the $SU(3)$ symmetry breaking effects.
The probabilities of light quark-antiquark pairs for each of the ground-state baryons in terms of the
relevant configurations ($P_B (i)$) are given below.

\begin{itemize}
  \item {\bf Proton:}
\end{itemize}	
\vspace{-1.3cm}
\begin{eqnarray}
\mathcal{P}_p^{u \bar u} &=& [CG]_{23} \Big{[}P_p(3)+P_p(5)+P_p(9)+P_p(12)+P_p(16)\Big{]} , \\
	\mathcal{P}_p^{d \bar d} &=& [CG]_{13} \Big{[}P_p(3)+P_p(5)+P_p(9)+P_p(12)+P_p(16)\Big{]}	\nonumber \\
	               &+& [CG]_{11} \Big{[}P_p(1)+P_p(8)+P_p(15)\Big{]} .
\end{eqnarray}
\begin{itemize}
  \item {\bf Neutron:}
\end{itemize}	
\vspace{-1.3cm}	
\begin{equation}
	\mathcal{P}_n^{u \bar u} = \mathcal{P}_p^{d \bar d}	~;~	\mathcal{P}_n^{d \bar d} = \mathcal{P}_p^{u \bar u}.
\end{equation}
\begin{itemize}
  \item {$\bf \Lambda :$}
\end{itemize}		
\vspace{-1.3cm}
\begin{eqnarray}
\mathcal{P}_\Lambda^{u \bar u}&=& [CG]_{12} \Big{[}P_\Lambda(1) + P_\Lambda(2) + P_\Lambda(4) + \mathcal{P}_\Lambda(6)+P_\Lambda(7)+ P_\Lambda(8)  \nonumber\\
	               &+& P_\Lambda(10)+P_\Lambda(11) + P_\Lambda(13) + P_\Lambda(14) + P_\Lambda(15) +P_\Lambda(17)\Big{]} , \\
\mathcal{P}_\Lambda^{d \bar d} &=& \mathcal{P}_\Lambda^{u \bar u} .
\end{eqnarray}
\begin{itemize}
  \item {$\bf \Sigma^+ :$}
\end{itemize}	
\vspace{-1.3cm}
\begin{eqnarray}
\mathcal{P}_{\Sigma^+}^{u \bar u} &=& [CG]_{34} \Big{[} {P}_{\Sigma}(3)+P_{\Sigma}(5)+P_{\Sigma}(9)+P_{\Sigma}(12)+P_{\Sigma}(16)\Big{]} , \\	
\mathcal{P}_{\Sigma^+}^{d \bar d} &=& [CG]_{14} \Big{[} P_{\Sigma}(3) + P_{\Sigma}(5) +P_{\Sigma}(9)+ P_{\Sigma}(12)+ {P}_{\Sigma}(16) \Big{]} \nonumber\\
                            &+& [CG]_{11} \Big{[} P_{\Sigma}(1) +P_{\Sigma}(2)+ P_{\Sigma}(4) +P_{\Sigma}(6)+ {P}_{\Sigma}(7) + P_{\Sigma}(8)\nonumber\\
                            &+& P_{\Sigma}(10) +P_{\Sigma}(11)+ P_{\Sigma}(13) +P_{\Sigma}(14)+ P_{\Sigma}(15) + {P}_{\Sigma}(17)\Big{]} .
\end{eqnarray}
\begin{itemize}
  \item {$\bf \Sigma^\circ :$}
\end{itemize}	
\vspace{-1.3cm}
\begin{eqnarray}
\mathcal{P}_{\Sigma^\circ}^{u \bar u} &=& \mathcal{P}_{\Sigma^\circ}^{d \bar d}
= [CG]_{12} \sum_{i=1}^{17} P_{\Sigma}(i)   \\
      &=& \frac{1}{2}(\mathcal{P}_{\Sigma^+}^{u \bar u} + \mathcal{P}_{\Sigma^+}^{d \bar d}).
\end{eqnarray}
\begin{itemize}
  \item {$\bf \Sigma^- :$}
\end{itemize} 	
\vspace{-1.3cm}
\begin{equation}
\mathcal{P}_{\Sigma^-}^{u \bar u} = \mathcal{P}_{\Sigma^+}^{d \bar d} ~;~    	
\mathcal{P}_{\Sigma^-}^{d \bar d} = \mathcal{P}_{\Sigma^+}^{u \bar u} .
\end{equation}
\begin{itemize}
  \item {$\bf \Xi^\circ :$}
\end{itemize}	
\vspace{-1.3cm}
\begin{eqnarray}
\mathcal{P}_{\Xi^\circ}^{u \bar u} &=& [CG]_{23}
	\Big{[}P_{\Xi}(1)+P_{\Xi}(3)+P_{\Xi}(5)+P_{\Xi}(9)+P_{\Xi}(12)+P_{\Xi}(15)+ P_{\Xi}(16) \Big{]} \nonumber\\
                &+&[CG]_{11} \Big{[}P_{\Xi}(6)+P_{\Xi}(8)\Big{]} ,  \\
\mathcal{P}_{\Xi^\circ}^{d \bar d} &=& [CG]_{13} \Big{[}P_{\Xi}(1)+P_{\Xi}(3)+P_{\Xi}(5)+P_{\Xi}(9)+P_{\Xi}(12)+P_{\Xi}(15)+ P_{\Xi}(16) \Big{]} \nonumber\\
	                &+&[CG]_{11} \Big{[}P_{\Xi}(2) + P_{\Xi}(4) +P_{\Xi}(7)+ P_{\Xi}(10) +P_{\Xi}(11)+ P_{\Xi}(13) +P_{\Xi}(14)\nonumber\\
	                &+& P_{\Xi}(17)\Big{]} .
\end{eqnarray}
\begin{itemize}
  \item {$\bf \Xi^- :$}
\end{itemize} 	
\vspace{-1.3cm}
\begin{equation}
\mathcal{P}_{\Xi^-}^{u \bar u} = \mathcal{P}_{\Xi^\circ}^{d \bar d} ~;~
\mathcal{P}_{\Xi^-}^{d \bar d} = \mathcal{P}_{\Xi^\circ}^{u \bar u} .
\end{equation}

For the $s \bar {s}$ component, the probabilities $\mathcal{P}_{B}^{s \bar s}$ are obtained by
summing up linearly the relevant nonvanishing contributions, ${P}_{B}^{s \bar s} (i)$ (i=1,17).
%
\subsection{Sigma terms}
\label{sec:s-t}
Having obtained the required ingredients, we focus on the pion-baryon and the strangeness-baryon
$\sigma$ terms as a function of the quark-antiquark pairs probabilities.


The pion-nucleon and strangeness-nucleon $\sigma$ terms are defined as follows:
\begin{eqnarray}
 \sigma_{\pi N} & = & m_{l}\langle N|u \bar{u}+d \bar{d}|N\rangle\,, \\
 \sigma_{s N} & = & m_{s} \langle N|s \bar{s}|N\rangle ,
\end{eqnarray}
with $m_{l}\equiv(m_{u}+m_{d})/2$ the average current mass of the up and down quarks and
$m_{s}$ the current mass of the strange quark.

$\sigma_{\pi N}$ can be related to the nucleon expectation value of the purely
octect operator ($\hat{\sigma}$) and to the strangeness content of the nucleon
($y_{_N}$), respectively,
\begin{eqnarray}
\hat{\sigma}&=&m_{l}\langle N|u \bar{u}+d \bar{d}-2 s \bar{s}|N\rangle\,, \\ [10pt]
 y_{_N}&=&\frac{2 \langle N|s \bar{s}|N\rangle}{\langle N|u \bar{u} +d \bar{d}|N\rangle} .
\end{eqnarray}

Hence,
\begin{eqnarray}
 \label{piNy}
\sigma_{\pi N} &=& \frac{\hat{\sigma}}{1- y_{_N}}\,, \\ [10pt]
\sigma_{s N} &=& \frac{m_s}{m_l} y_{_N} \sigma_{\pi N}.
\label{SN}
\end{eqnarray}

In the following we give explicit expressions for $\sigma_{\pi B}$ and $\sigma_{s B}$
in terms of the five-quark probabilities $\mathcal{P}_B^{Q \bar Q}$.
%
\begin{itemize}
  \item {\bf Nucleon:}
\end{itemize}	
\vspace{-1.4cm}
\begin{eqnarray}
\label{piN}
\sigma _{\pi N} &=&  \frac{3+2(\mathcal{P}_N^{u \bar u}+\mathcal{P}_N^{d \bar d})}
     {3+2(\mathcal{P}_N^{u \bar u}+\mathcal{P}_N^{d \bar d}-2\mathcal{P}_N^{s \bar s})} \hat{\sigma}, \\ [10pt]
y_{_N} &=&
  \frac{2\mathcal{P}_N^{s \bar s}}{3+2(\mathcal{P}_N^{u \bar u}+\mathcal{P}_N^{d \bar d})}.
\label{yN}
\end{eqnarray}
%
\begin{itemize}
  \item {\bf {S=-1 hyperons ($Y \equiv \Lambda,~\Sigma$):}}
\end{itemize}	
\vspace{-1.4cm}
\begin{eqnarray}
 \label{piY}
\sigma_{\pi Y} &=&  \frac{2+ 2 (\mathcal{P}_Y^{u \bar u}+\mathcal{P}_Y^{d \bar d})}
        {3+2(\mathcal{P}_N^{u \bar u}+\mathcal{P}_N^{d \bar d})} \sigma_{\pi N} , \\ [10pt]
\sigma_{s Y} &=&  \frac{1+2\mathcal{P}_Y^{s \bar s}}
          {2 \mathcal{P}_N^{s \bar s}} \sigma_{s N}\nonumber  \\ [10pt]
\label{SY1}
          &=&
          \frac{m_s}{m_l} \frac{1+2\mathcal{P}_Y^{s \bar s}}
          {3+2(\mathcal{P}_N^{u \bar u}+\mathcal{P}_N^{d \bar d})} \sigma_{\pi N}.
\label{SY2}
\end{eqnarray}
%
\begin{itemize}
  \item {\bf {S=-2 hyperons $\Xi$:}}
\end{itemize}	
\vspace{-1.4cm}
\begin{eqnarray}
\sigma_{\pi \Xi} &=& \frac{1+ 2 (\mathcal{P}_\Xi^{u \bar u}+\mathcal{P}_\Xi^{d \bar d})}
        {3+2(\mathcal{P}_N^{u \bar u}+\mathcal{P}_N^{d \bar d})} \sigma_{\pi N}, \\ [10pt]
 \label{piXi}
\sigma_{s \Xi} &=& \frac{2+ 2 \mathcal{P}_\Xi^{s \bar s}}
          {2 \mathcal{P}_N^{s \bar s}} \sigma_{s N}\nonumber  \\ [10pt]
\label{SXi1}
          &=&
          \frac{m_s}{m_l} \frac{2+ 2 \mathcal{P}_\Xi^{s \bar s}}
          {3+2(\mathcal{P}_N^{u \bar u}+\mathcal{P}_N^{d \bar d})} \sigma_{\pi N}.
\label{SXi2}
\end{eqnarray}
%
%
\section{Results and Discussion}
\label{sec:Res}
The probabilities per configuration, introduced in Sec.~\ref{sec:proba} and intervening in
Eqs.~(\ref{piN}) - (\ref{SXi2}), are reported in Appendix \ref{A-proba}, Tables~\ref{A1} and~\ref{A2}.
Using values for the probabilities in the latter Tables and equations in Secs. ~\ref{sec:proba}
and~\ref{sec:s-t}, we calculated the $\sigma$ terms per configuration and per baryon.

Before moving to the presentation of our results a comment on the uncertainties is
in order.
The parameters of our extended constituent quark model are reported in Ref.~\cite{An:2012kj}.
The only source of uncertainty in the probabilities (Tables~\ref{A1} and~\ref{A2}) comes from a
common factor of the matrix elements of the transitions between three- and five-quark components and
was found~\cite{An:2012kj} to be $V$=570$\pm$46 MeV, by successfully fitting the experimental data
for the proton flavor asymmetry $\bar{d} - \bar{u}$.
For the $\sigma$ terms two additional entities contribute to the uncertainties, namely,
the nonsinglet component $\hat{\sigma}=33 \pm 5$~MeV as extracted within the chiral
perturbation theory~\cite{Borasoy:1996bx} and the PDG masses ratio~\cite{Agashe:2014kda}
$m_{s}/m_{l} = 27.5 \pm 1.0$.
Accordingly, no parameters were adjusted in the frame of the present work.

%
\begin{table}[t]
\caption{\footnotesize Predictions for the nucleon strangeness content parameter $y_{_N}$,
$\sigma_{\pi N}$ and $\sigma_{s N}$ (MeV), per configuration i=1, 17 and per category I to IV.
\label{st-N}}
\scriptsize
\begin{tabular}{rccccc}
\hline\hline
i & Category  & $y_{_N}$ && $\sigma_{\pi N}$ & $\sigma_{sN}$ \\
\hline
  & ~~~~I) $[31]_{X}[22]_{S}$: & & & \\
 1 &              & 0.006 (1) && 33.4 (5.1) & 5.5 (1.5) \\
 2 &              & 0.002 (0) && 33.2 (5.0) & 2.2 (0.6) \\
 3 &              & 0         && 33.0 (5.0) & 0 \\
 4 &              & 0.002 (0) && 33.1 (5.0) & 1.6 (0.5) \\
   & ~Category~I  & 0.010 (1) && 33.6 (5.2) & 8.9 (2.5) \\ [10pt]
%
%
  & ~~~~II) $[31]_{X}[31]_{S}$: & && & \\
 5 &               & 0         && 33.0 (5.0) & 0 \\
 6 &               & 0.004 (0) && 33.3 (5.1) & 3.8 (1.1) \\
 7 &               & 0.002 (0) && 33.1 (5.0) & 2.0 (0.6) \\
 8 &               & 0.001 (0) && 33.1 (5.0) & 1.1 (0.3) \\
 9 &               & 0         && 33.0 (5.0) & 0 \\
 10 &              & 0.001 (0) && 33.0 (5.0) & 0.5 (0.1) \\
    & ~Category~II & 0.008 (1) && 33.5 (5.1) & 7.0 (2.0) \\ [10pt]
%
%
  & ~~~~III) $[4]_{X}[22]_{S}$: && & & \\
 11 &              & 0.006 (1) && 33.4 (5.1) & 5.2 (1.5) \\
 12 &              & 0         && 33.0 (5.0) & 0 \\
 13 &              & 0.004 (0) && 33.3 (5.1) & 4.0 (1.2) \\
   & ~Category III & 0.010 (1) && 33.7 (5.2) & 9.0 (2.6) \\ [10pt]
%
%
  & ~~~~IV) $[4]_{X}[31]_{S}$: && & & \\
 14 &              & 0.005 (1) && 33.3 (5.1) & 4.7 (1.4) \\
 15 &              & 0.003 (0) && 33.2 (5.0) & 2.6 (0.8) \\
 16 &              & 0         && 33.0 (5.0) & 0 \\
 17 &              & 0.001 (0) && 33.1 (5.0) & 1.2 (0.3) \\
   & ~Category~IV  &  0.009 (1)&& 33.6 (5.2) & 8.5 (2.5) \\  [10pt]
   & All configurations & 0.031(3) && 35.2 (5.5) & 30.5 (8.5) \\
\hline
\hline
\end{tabular}
\end{table}
%
%

Detailed results for the nucleon are given in Table~\ref{st-N}.
From Eq.~(\ref{yN}) it can be inferred that the strangeness content parameter $y_{_N}$ per
configuration (Table~\ref{st-N}, column 3)
is $\approx 2/3 P_B^{s \bar s} (i)$, given that for any configuration,
$2({P}_N^{u \bar u}+{P}_N^{d \bar d})\ll 3$ (Table~\ref{A1}, column 6).
That approximation almost holds also for the results per category, but breaks at the level of 25\%
for the full calculation embodying all 17 configurations (Table~\ref{st-N}, last row
of column 3), due to the (sizable) probabilities of five-quark components in the nucleon
(Table~\ref{A1}, last row); $\mathcal{P}_N^{u \bar u}+\mathcal{P}_N^{d \bar d}=0.313$ and
$\mathcal{P}_N^{s \bar s}=0.058$.
It is worth noting that out of the 17 five-quark configurations in the nucleon,
only 3 of them contribute to both light and $s \bar {s}$ pair probabilities, whereas 5 of
them have only $u \bar {u}$ and  $d \bar {d}$ components, while the remaining 9 configurations
are exclusively composed of $s \bar {s}$ pairs; see Table~\ref{A1}.
Accordingly, any configuration truncated set would alter not only the respective probabilities
in light and / or strange sectors, but also would change their relative probabilities and, hence,
$y_{_N}$.

The strangeness-nucleon sigma term being proportional to $y_{_N}$ [Eq.~\ref{SN}],
$\sigma_{sN}$ shows similar behaviors with respect to the probabilities of light
and strange quark-antiquark pairs (Table~\ref{st-N}, last column).
Moreover, while $\sigma_{sN}$ per configuration stays between 7 and 9 MeV, the complete
calculation with all 17 configurations leads to a value more than 3 times larger
(Table~\ref{st-N}, last row).

{\it The above two paragraphs allow establishing that any configuration truncated
model will significantly underestimate both $y_{_N}$ and $\sigma_{sN}$.}
A similar conclusion was reached in our study~\cite{An:2013daa} on the strangeness magnetic
form factor of the proton, where the truncation effects were found to be even more drastic.

Given that the nucleon structure is dominated by the three-light-quark component, 
the pion-nucleon $\sigma$ term shows much less sensitivity to the probabilities, 
leading to results per configuration or per category close to the final value (Table~\ref{st-N},
column 4).
Actually, Eq.~(\ref{piN}) puts $\sigma _{\pi N} \approx \hat{\sigma}$.
Note that the nucleon expectation value of the purely octet operator
$\hat{\sigma}$ is still under debate~\cite{Alarcon:2012nr}.

Hyperon $\sigma$ terms are related to $\sigma _{\pi N}$ [Eqs.~(\ref{piY}] to~(\ref{SXi2}),
and taking $m_s / m_l \approx$27, one obtains the following approximations, where
$Y \equiv  \Lambda,~\Sigma$:
\begin{equation}
 \sigma_{\pi Y} \approx (2/3)\sigma_{\pi N},~
 \sigma_{s Y} \approx 9\sigma_{\pi N},~
 \sigma_{\pi \Xi} \approx (1/3)\sigma_{\pi N},~
 \sigma_{s \Xi} \approx 18\sigma_{\pi N}.
 \label{Sappro}
\end{equation}

Accordingly, the hyperon sigma terms vary slightly from one configuration / category
to another, as was discussed for $\sigma_{\pi N}$,
so we do not show detailed results per configuration for hyperons, but only summarize the
numerical results per category in Table~\ref{cats}.
%
\begin{table}[htb]
\begin{center}
\caption{\footnotesize Predictions per category for the $\sigma$ terms of the octet baryons (MeV).}
\label{cats}
\begin{tabular}{ccccccccccccccc}
\hline\hline
&& \multicolumn{2}{c}{N} && \multicolumn{2}{c}{$\Lambda$} && \multicolumn{2}{c}{$\Sigma$}
&& \multicolumn{2}{c}{$\Xi$}\\
\hline
 Category
&&  ${\sigma}_{\pi N}$ & ${\sigma}_{s N}$
&&  ${\sigma}_{\pi \Lambda}$ & ${\sigma}_{s \Lambda}$
&&${\sigma}_{\pi \Sigma}$ & ${\sigma}_{s \Sigma}$
&&${\sigma}_{\pi \Xi}$ & ${\sigma}_{s \Xi}$
\\
\hline
I  && 33.6$\pm$5.2 & 8.9$\pm$2.5 && 22.9$\pm$3.6 & 283$\pm$51 && 22.1$\pm$3.4 & 292$\pm$54
 && 12.2$\pm$2.0 & 562$\pm$101   \\
II && 33.5$\pm$5.1 & 7.0$\pm$2.0 && 22.6$\pm$3.5 & 301$\pm$56 && 22.8$\pm$3.5 & 299$\pm$56
 && 11.8$\pm$1.9 & 582$\pm$109  \\
III&& 33.7$\pm$5.2 & 9.0$\pm$2.6 && 22.7$\pm$3.5 & 313$\pm$60 && 23.3$\pm$3.4 & 307$\pm$58
 && 12.0$\pm$1.9 & 614$\pm$116  \\
IV && 33.6$\pm$5.2 & 8.5$\pm$2.5 && 22.9$\pm$3.6 & 309$\pm$59 && 22.5$\pm$3.5 & 311$\pm$59
 && 12.2$\pm$2.0 & 609$\pm$115  \\
All  && 35.2$\pm$5.5 & 30.5$\pm$8.5&& 25.0$\pm$4.1 & 297$\pm$55 && 24.5$\pm$4.0 & 301$\pm$57
 && 14.8$\pm$2.6 & 564$\pm$102  \\
\hline
\hline
\end{tabular}
\end{center}
\end{table}
%
%

The pion-hyperon $\sigma$ terms per category turn out to be smaller than the final results,
while the strangeness-baryon terms oscillate around the final value, with deviations staying
within the uncertainties.

In Table~\ref{compa}, we compare our results with recent findings by other authors.
Note that, wherever appropriate, using statistical and systematic uncertainties
reported in those papers, we give $\delta = \sqrt{\delta^2_{stat} + \delta^2_{sys}}$.
%
%
{\squeezetable
\begin{table}[htb]
\begin{center}
\caption{\footnotesize Predictions for the $\sigma_{\pi B}$ and $\sigma_{sB}$ of the octet
baryons (MeV).}
\label{compa}
\scriptsize
\begin{tabular}{lccccccccccccccccc}
\hline\hline
%
 Reference (Collaboration)&  Approach
&& $y_{_N}$ && ${\sigma}_{\pi N}$ & ${\sigma}_{s N}$
&&  ${\sigma}_{\pi \Lambda}$ & ${\sigma}_{s \Lambda}$
&&${\sigma}_{\pi \Sigma}$ & ${\sigma}_{s \Sigma}$
&&${\sigma}_{\pi \Xi}$ & ${\sigma}_{s \Xi}$
\\
\hline
Present work & E$\chi$CQM && 0.03$\pm$0.00 && 35$\pm$5 & 30$\pm$8&& 25$\pm$4 & 297$\pm$55 && 24$\pm$4 & 301$\pm$57
&& 15$\pm$3 & 564$\pm$102  \\
Semke-Lutz~\cite{Semke:2012gs} & $\chi$Lagrangian && 0.05$\pm$0.04 && 32$\pm$2 & 22$\pm$20 && 22$\pm$2 & 214$\pm$24 &&
 15$\pm$2 & 292$\pm$19 && 9$\pm$2 & 414$\pm$35  \\
Lutz {\it et al.}~\cite{Lutz:2014oxa} & $\chi$Lagrangian &&  && 39$\pm$1 & 84$\pm$3 && 23$\pm$1 & 230$\pm$3 &&
 18$\pm$1 & 355$\pm$5 && 6$\pm$1 & 368$\pm$8  \\
Ren {\it et al.}~\cite{Ren:2013dzt} & $\chi$PT && 0.24$\pm$0.12 && 43$\pm$6 & 126$\pm$59 && 19$\pm$7 & 269$\pm$70 &&
18$\pm$6 & 296$\pm$54 && 4$\pm$3 & 397$\pm$60 \\
Ren {\it et al.}~\cite{Ren:2014vea} & $\chi$PT && * &&
 55$\pm$1 & 27$\pm$27 && 32$\pm$2 & 185$\pm$29 && 34$\pm$3 & 210$\pm$49
 && 16$\pm$2 & 333$\pm$28 \\
Alarcon {\it et al.}~\cite{Alarcon:2012nr} & $\chi$PT && 0.02$\pm$0.16 && 59$\pm$7 & 16$\pm$100 && &&  &  &&  &  \\
Bali {\it et al.}~\cite{Bali:2011ks} (QCDSF) & LQCD && $<$ 0.14 && 38$\pm$12 & 12$\pm$28 &&  &&  &  &&  &   \\
Durr {\it et al.}~\cite{Durr:2011mp} (BMW) & LQCD && 0.20$\pm$0.17 && 39$\pm$13 & 67$\pm$58 && 29$\pm$8 & 180$\pm$68 &&
28$\pm$11 & 245$\pm$68 && 16$\pm$6 & 312$\pm$82  \\
Horsley {\it et al.}~\cite{Horsley:2011wr} (QCDSF-UKQCD) & LQCD && 0.18$\pm$0.17 &&
 31$\pm$5 & 71$\pm$68 && 24$\pm$5 & 247$\pm$77 && 21$\pm$5 & 336$\pm$77 && 16$\pm$5 & 468$\pm$69 \\
Alexandrou {\it et al.}~\cite{Alexandrou:2013uwa} (ETM) & LQCD && 0.13$\pm$0.05 && 37$\pm$25 & 28$\pm$13 &&  && & && & \\
\hline
\hline
* See the text.
\end{tabular}
\end{center}
\end{table}
}

Semke and Lutz~\cite{Semke:2012gs} determined the sigma terms via a relativistic chiral
Lagrangian and large-N$_c$ sum rules at N$^3$LO, using different LQCD data as input.
That work also led to the small value for $y_{_N}$ and to $\sigma$ terms compatible with our
results, though within 2$\sigma$ for the largest discrepancies, i.e.,
${\sigma}_{s \Lambda}$, ${\sigma}_{\pi \Sigma}$, and ${\sigma}_{\pi \Xi}$.
By considering the finite volume effects, Lutz and collaborators~\cite{Lutz:2014oxa}
improved significantly the relativistic chiral Lagrangian approach.
The 12 free parameters of the model were successfully extracted by fitting about 220 data
points reported by six different LQCD groups. It is worth noting that a full estimate of
the systematic uncertainties in that approach requires investigating the corrections to the
large-N$_c$ sum rules; a task not yet performed. Hence, the uncertainties turn out to
be rather small (Table~\ref{compa}). The most striking feature due to the finite volume effects    
is the large value obtained for ${\sigma}_{s N}$, which falls in the large ranges 
determined by BMW and QCDSF-UKQCD groups, but is not consistent with findings from other
predictions quoted in Table~\ref{compa}.

Ren and collaborators~\cite{Ren:2013dzt,Ren:2014vea} performed studies
within the extended-on-mass-shell renormalization scheme in the baryon chiral
perturbation theory up to N$^3$LO.
The earlier work~\cite{Ren:2013dzt} led to $y_{_N}$=0.24$\pm$0.12,
${\sigma}_{\pi N}$=43$\pm$6 MeV and ${\sigma}_{s N}$=126$\pm$59 MeV.
In the latest work~\cite{Ren:2014vea}, adopting a more selective and restricted LQCD
data set, ${\sigma}_{\pi N}$ increases slightly, while ${\sigma}_{s N}$ drops down to
27$\pm$27 and, hence, $y_{_N}$ gets reduced by roughly a factor of 5,
close enough to our value.

Alarcon and collaborators in a recent calculation~\cite{Alarcon:2012nr}, based on the Lorentz covariant
chiral perturbation theory at $\mathcal {O}(p^3)$ taking $\sigma_{\pi N}$=59$\pm$7 MeV
~\cite{Alarcon:2011zs}, obtained $y_{_N}=$0.02$\pm$0.16 and $\sigma_{s N}$=16$\pm$100 MeV,
both consistent with our results, albeit with large uncertainties.

Lattice quantum chromodynamics (LQCD) approaches have put forward predictions for the
$\sigma$ terms.
Here, we limit ourselves to a few recent ones; rather exhaustive references can be found in,
e.g., Refs.~\cite{Alarcon:2012nr,Ren:2014vea}.
Bali and collaborators~\cite{Bali:2011ks}, studying nucleon mass data with quark flavors $N_f=2$,
put an upper limit on $y_{_N}$ ($<$0.14).
Durr~\cite{Durr:2011mp} and Horsley~\cite{Horsley:2011wr} and collaborators extended
the study to $N_f=2+1$, and reported results for $y_{_N}$ and ${\sigma}_{s N}$ with large
uncertainties.
Alexandrou and collaborators~\cite{Alexandrou:2013nda,Alexandrou:2013uwa} performed
calculations using $N_f$=2+1+1 flavors of maximally twisted mass fermions.
Though the $y_{_N}$ is rather large, ${\sigma}_{s N}$ turns out to endorse our value.

A general trend in the LQCD calculation is that the systematic errors are substantially larger
than the statistical one(s).
Detailed investigations of the systematic errors in the
computation of the sigma terms and $y_{_N}$ were reported in
Refs.~\cite{Alexandrou:2013nda,Alexandrou:2013uwa}, quantifying contributions from various
sources, namely, the chiral extrapolation, the excited states contamination and the
discretization.
Note that therein, $y_{_N}$ was obtained directly by a ratio of three point function, while other
LQCD works for $N_f$=2+1 provide indirect determination of that quantity ~\cite{Shanahan:2012wh,Durr:2011mp,Horsley:2011wr}.
As reported in Ref.~\cite{Alexandrou:2013nda}, computing directly the ratio of the matrix
element allows avoiding any assumptions on the domain of validity of effective field theory
relations which is based on SU(2) or SU(3) chiral perturbation expansion and sometimes known
only at leading order accuracy.
Figure 2 in Ref.~\cite{Alexandrou:2013nda} is instructive regarding the interplay among
$y_{_N}$, $\sigma_{\pi N}$, $\sigma_{s N}$ and $\hat{\sigma}$.

Finally, with respect to the hyperon $\sigma$ terms, predictions from various approaches
(columns 6 - 11 in Table~\ref{compa}) result in values compatible with each other, albeit with
rather large uncertainties on the strangeness-hyperon terms (except for the chiral
Lagrangian~\cite{Semke:2012gs,Lutz:2014oxa} findings).
%

\section {Summary and conclusions}
\label{sec:conclu}
In the present work, we investigated the $\sigma$ terms of the ground-state octet baryons,
employing the recently developed extended chiral constituent quark model, within which the
baryons are considered as admixtures of three- and five-quark states.
Probabilities of the five-quark components were calculated using the $^{3}$P$_{0}$ transition
operator. The set of adjustable parameters in our approach were given in a previous
work~\cite{An:2012kj} dedicated to the flavor asymmetry of the nucleon sea $\bar{d}-\bar{u}$.
That set was used in a subsequent paper~\cite{An:2013daa} and allowed us to predicting
successfully
the strangeness magnetic form factor of the proton. In the present work we kept with the same
set of parameters to put forward predictions for the ground-state baryon octet sigma terms.

The determination of the strangeness content of the nucleon, the $y_{_N}$ parameter, the extraction
of the strangeness-nucleon sigma term, ${\sigma}_{s N}$, and of the ${\sigma}_{\pi N}$ are among
the crucial issues in this realm, going beyond the hadron physics and were found to be important in
topics such as dark matter in WIMP-nucleon~\cite{Alexandrou:2013uwa,DM} or
neutralino-nucleon~\cite{neutralino} cross sections, and in neutron matter~\cite{Kruger:2013iza}.

Our full calculation gives $y_{_N}$=0.031$\pm$0.003, ${\sigma}_{s N}$=30$\pm$8 MeV and
${\sigma}_{\pi N}$=35$\pm$5 MeV. So, the small value for $y_{_N}$ fulfills partially the OZI
rule, in agreement with results from the chiral Lagrangian~\cite{Semke:2012gs,Lutz:2014oxa} and 
chiral perturbation approaches~\cite{Ren:2014vea,Alarcon:2012nr}.
LQCD calculations~\cite{Bali:2011ks,Durr:2011mp,Horsley:2011wr,Alexandrou:2013uwa}
give larger central values but with roughly 40\% to 100\% uncertainties.
For ${\sigma}_{s N}$ our result is consistent with values reported within chiral
Lagrangian~\cite{Semke:2012gs} and the LQCD from the ETM Collaboration~\cite{Alexandrou:2013uwa}.
Other LQCD works discussed in the present paper led to too large uncertainties for significant
comparisons.
It is worth noting that our value for ${\sigma}_{s N}$ is very close to that determined in
Ref.~\cite{Gubler:2014pta} as  the value at which the $\phi$-meson mass-peak shift in nuclear
matter would undergo a sign change.
Accordingly, the expected accurate enough data~\cite{phi} on that effect would put significant
constraints on ${\sigma}_{s N}$.
Finally for ${\sigma}_{\pi N}$, our finding is compatible with various calculations based on
the chiral Lagrangian~\cite{Semke:2012gs,Lutz:2014oxa}, $\chi$PT~\cite{Ren:2013dzt}
and LQCD~\cite{Bali:2011ks,Durr:2011mp,Horsley:2011wr,Alexandrou:2013uwa}.

Within 2$\sigma$, the pion-hyperon sigma term (${\sigma}_{\pi \Lambda}$,
${\sigma}_{\pi \Sigma}$,
${\sigma}_{\pi \Xi}$) predictions from all the approaches quoted in Table~\ref{compa} come out
in comparable ranges.
This is not fully the case for ${\sigma}_{s \Lambda}$, ${\sigma}_{s \Sigma}$, and
${\sigma}_{s \Xi}$; the main disagreements with our values being with results coming from a
$\chi$PT~\cite{Ren:2013dzt} and (partially) with a LQCD approach~\cite{Durr:2011mp}.

In summary, the general trend is a convergence among predictions by different
approaches, within the reported uncertainties.
However, to achieve conclusive numerical results several challenging issues deserve 
further investigation, such as the large number of unknown
low-energy constants appearing at $\mathcal {O}(p^4)$ in the $\chi$PT approaches~\cite{Alarcon:2012nr},
dominant systematic errors in the LQCD calculations~\cite{Torrero:2014pxa}, and
the chiral Lagrangian approaches~\cite{Semke:2012gs,Lutz:2014oxa}, as well as the separation of
contributions from the three-quark and five-quark components in the baryons within the chiral
constituent quark formalism.

Finally, we emphasize the sizable sensitivities of $y_{_N}$ and ${\sigma}_{s N}$ to the
five-quark configurations set in our approach, showing that any truncated set will lead to
significant deviations from the predictions obtained by the full calculation embodying all 17
five-quark configurations.
%
\begin{acknowledgments}
This work is partly supported by the National Natural Science Foundation of China under Grant
No. 11205164 and the Fundamental Research Funds for the Central Universities under Grants
 No. XDJK2015C150 and No. SWU115020.
\end{acknowledgments}
%
%
%
\newpage
%
%
\begin{appendix}

\section{Probabilities of the quark-antiquark components in the ground-state octet baryons}
\label{A-proba}

\begin{table}[hbt]
\caption{\footnotesize Predictions for probabilities of different five-quark
configurations for the nucleon and $\Lambda$ (in \%), with
$\mathcal{P}^{q\bar{q}}_B = \mathcal{P}^{u\bar{u}}_B + \mathcal{P}^{d\bar{d}}_B$,
$\mathcal{P}^{Q\bar{Q}}_B = \mathcal{P}^{q\bar{q}}_B + \mathcal{P}^{s\bar{s}}_B$.
\label{A1}}
\scriptsize
\begin{tabular}{rcccccccccc}
\hline\hline
i & Category  & $P^{u\bar{u}}_p$ & $ P^{d\bar{d}}_p$ & $P^{q\bar{q}}_N$ & $P^{s\bar{s}}_N$ & $P^{Q\bar{Q}}_N$ && $P^{q\bar{q}}_\Lambda$ & $P^{s\bar{s}}_\Lambda$ & $P^{Q\bar{Q}}_\Lambda$ \\
%
%
\hline \\
  & ~~~~I) $[31]_{X}[22]_{S}$: & &&&& && &  &   \\
 1 &  & 0 & 14.6 (1.5) & 14.6 (1.5) & 1.0 (0.1) & 15.6 (1.5) && 11.3 (1.2) & 0 & 11.3 (1.2) \\
 2 &  & 0 & 0          & 0          & 0.3 (0.1) & 0.3 (0.1)  && 0.3 (0.0) & 0.3 (1) & 0.6 (0.1) \\
 3 &  & 1.1 (0.1) & 0.5 (0.1) & 1.6 (0.2) & 0 &1.6 (0.2) && 0 & 0 & 0 \\
 4 &  & 0 & 0 & 0 & 0.3 (0.1) & 0.3 (0.1) && 1.4 (0.1) & 0.6 (0.1) & 2.0 (0.3) \\
   & ~Category~I   & 1.1 (0.1) & 15.1 (1.5) & 16.2 (1.6) & 1.7 (0.2) & 17.9 (0.8)  && 13.0 (1.4) & 0.9 (0.2) & 13.9 (1.6) \\ \\
%
%
  & ~~~~II) $[31]_{X}[31]_{S}$: & &&&& && &  & \\
 5 &  & 4.8 (0.5) & 2.4 (0.2) &  7.2 (0.7) & 0 & 7.2 (0.7)  && 0 & 0 & 0 \\
 6 &  & 0 & 0 & 0 & 0.6 (0.1) & 0.6 (0.1)   && 5.1 (0.5) & 1.3 (0.2) & 6.4 (0.7) \\
 7 &  & 0 & 0 & 0 & 0.3 (0.1) & 0.3 (0.1)   && 0.3 (0.0) & 0.2 (0.1) & 0.5 (0.1) \\
 8 &  & 0 & 0.6 (0.1) & 0.6 (0.1) & 0.2 (0.0) & 0.8 (0.1)  && 1.0 (0.1) & 0 & 1.0 (0.1) \\
 9 &  & 0.3 (0.0) & 0.2 (0.0) &  0.5 (0.1) & 0 & 0.5 (0.1) && 0 & 0 & 0 \\
 10 & & 0 & 0 & 0 & 0.1 (0.0) & 0.1 (0.0) && 0.4 (0) & 0.2 (0) & 0.6 (0.1) \\
   & ~Category~II & 5.1 (0.5) & 3.2 (0.3)  & 8.3 (0.8) & 1.2 (0.1) & 9.5 (1.0) && 6.8 (0.6) & 1.7 (0.2) & 8.5 (0.9)  \\ \\
%
%
  & ~~~~III) $[4]_{X}[22]_{S}$: & &&&& && &  & \\
 11 &  & 0 & 0 & 0 & 0.9 (0.1) & 0.9 (0.1)  && 0.6 (0.1) & 0.6 (0.1) & 1.2 (0.1) \\
 12 &  & 2.7 (0.3) & 1.4 (0.1) &  4.1 (0.4) & 0 & 4.1 (0.4) && 0 & 0 & 0 \\
 13 &  & 0 & 0 & 0 & 0.7 (0.1) & 0.7 (0.1)   && 3.6 (0.4) & 1.5 (0.2) & 5.1 (0.6) \\
   & ~Category III &  2.7 (0.3) & 1.4 (0.1) &  4.1 (0.4) & 1.6 (0.2) & 5.7 (0.6)   && 4.2 (0.4)  & 2.1 (0.2) & 6.3(0.7) \\ \\
%
%
  & ~~~~IV) $[4]_{X}[31]_{S}$: & &&&& && &  & \\
 14 &  & 0 & 0 & 0 & 0.8 (0.1) & 0.8 (0.1)   && 1.5 (0.1) & 0.6 (0.1) & 2.1 (0.2) \\
 15 &  & 0 & 1.5 (0.2) & 1.5 (0.2) & 0.4 (0.1)& 1.9 (0.2)   && 2.4 (0.3)& 0 & 2.4 (0.3) \\
 16 &  & 0.8 (0.1) & 0.4 (0.0) &  1.2 (0.1) & 0 & 1.2 (0.1)   && 0 & 0 & 0 \\
 17 &  & 0 & 0 & 0 & 0.2 (0.0) & 0.2 (0.0)   && 1.0 (0.1) & 0.4 (0.1) & 1.4 (0.2) \\
   & ~Category~IV &  0.8 (0.1) & 1.9 (0.2) &  2.7 (0.3) & 1.4 (0.1) & 4.1 (0.4)   && 4.9 (0.5) & 1.0 (0.1) & 5.9 (0.6) \\ \\
   & All configurations & 9.7 (1.0) & 21.6 (2.2) &  31.3 (3.2) & 5.8 (0.6)& 37.1 (3.8)   && 28.9 (3.1) & 5.7 (0.6) & 34.6 (3.7) \\ \\
\hline
\hline
\end{tabular}
\end{table}
%
\begin{table}[hbt]
\caption{\footnotesize Same as Table~\ref{A1}, but for  $\Sigma$ and $\Xi$ hyperons (in \%).
\label{A2}}
\scriptsize
\begin{tabular}{rcccccccccccc}
\hline\hline
i & Category  & $P^{u\bar{u}}_{\Sigma^+}$ &
$ P^{d\bar{d}}_{\Sigma^+}$ &
$P^{q\bar{q}}_{\Sigma}$ & $P^{s\bar{s}}_{\Sigma}$ & $P^{Q\bar{Q}}_{\Sigma}$ &&
$P^{u\bar{u}}_{\Xi^\circ}$ & $P^{d\bar{d}}_{\Xi^\circ}$ & $P^{q\bar{q}}_{\Xi}$ & $P^{s\bar{s}}_{\Xi}$ & $P^{Q\bar{Q}}_{\Xi}$ \\
\hline \\
  & ~~~~I) $[31]_{X}[22]_{S}$: & &&&& && &  & &  &   \\
 1 &  & 0 & 6.7 (0.7) & 6.7 (0.7) & 2.0 (0.2) & 8.7 (1.0) && 5.4 (0.6) & 2.7 (0.3) & 8.0 (0.9) & 0 & 8.0 (0.9) \\
 2 &  & 0 & 1.0 (0.1) & 1.0 (0.1) & 0 & 1.0 (0.1)  && 0 & 0.9 (0.1) & 1.0 (0.1) & 0 & 1.0 (0.1) \\
 3 &  & 1.3 (0.1) & 0.4 (0.0) & 1.7 (0.2) & 0.4 (0.1) & 2.1 (0.2) && 0.7 (0.1) & 0.3 (0.0) & 1.0 (0.1) & 0.9 (0.1)  & 1.9 (0.2) \\
 4 &  & 0 & 0.1 (0) & 0.1 (0) & 0 & 0.1 (0.0) && 0 & 0.2 (0.0) & 0.2 (0.0) & 0 & 0.2 (0.0) \\
   & ~Category~I   & 1.3 (0.1) & 8.2 (0.9) & 9.5 (1.0) & 2.4 (0.3) & 11.9 (1.2)  && 6.1 (0.7)  & 4.1 (0.4)  & 10.2 (1.1) & 0.9 (0.1)  & 11.1 (1.1) \\ \\
%
%
  & ~~~~II) $[31]_{X}[31]_{S}$: & &&&& && &  & &  & \\
 5 &  & 4.1 (0.4) & 1.4 (0.1) &  5.5 (0.6) & 0.9 (0.1) & 6.4 (0.7)  && 1.8 (0.2) & 0.9 (0.1) & 2.7 (0.3) & 1.6 (0.2)  & 4.3 (0.5) \\
 6 &  & 0 & 0.3  (0.1) & 0.3 (0.1) & 0 & 0.3 (0.1)  && 0 & 0.6 (0.1) & 0.6 (0.1) & 0 & 0.6 (0.1) \\
 7 &  & 0 & 0.8  (0.1) & 0.8  (0.1) & 0 & 0.8 (0.1)   && 0 & 0.8 (0.1) & 0.8 (0.1) & 0  & 0.8 (0.1) \\
 8 &  & 0 & 0.4 (0.0) & 0.4 (0.0) & 0.5 (0.1) & 0.9 (0.0)  && 0.7 (0.1) & 0.4 (0.0) & 1.1 (0.1) & 0 & 1.1 (0.1) \\
 9 &  & 0.4 (0.0) & 0.1 (0.0) &  0.5 (0.1) & 0.1 (0) & 0.6 (0.1)&& 0.2 (0.0) & 0.1 (0.0) & 0.3 (0.1) & 0.3 (0.0)  & 0.6 (0.1) \\
 10 &  & 0 & 0.0 (0.0) & 0.0 (0.0) & 0 & 0.0 (0.0)   && 0 & 0.1 (0.0) & 0.1 (0.0) & 0 & 0.1 (0.0) \\
   & ~Category~II & 4.5 (0.3) & 3.0 (0.3)  & 7.5 (0.8) & 1.5 (0.1) & 9.0 (1.0) && 2.7 (0.3)  & 2.9 (0.3)  & 5.6 (0.6) & 1.9 (0.3)  & 7.5 (1.0)  \\ \\
%
%
  & ~~~~III) $[4]_{X}[22]_{S}$: & &&&& && &  & \\
 11 &  & 0 & 2.2  (0.2) & 2.2  (0.2) & 0 & 2.2 (0.2)  && 0 & 2.1 (0.2) & 2.1 (0.2) & 0 & 2.1 (0.2)   \\
 12 &  & 3.2 (0.4) & 1.1 (0.1) &  4.3 (0.5) & 1.1 (0.2) & 5.4 (0.6) && 1.7 (0.2) & 0.9 (0.1) & 2.6 (0.3) & 2.2 (0.2)  & 4.8 (0.5) \\
 13 &  & 0 & 0.2 (0) & 0.2 (0.0) & 0 & 0.2 (0.1)   && 0 & 0.5 (0.1) & 0.5 (0.1) & 0 & 0.5 (0.1) \\
   & ~Category III &  3.2 (0.4) & 3.5 (0.4) &  6.7 (0.7) & 1.1 (0.2) & 7.8 (0.7)   && 1.7 (0.2)  & 3.5 (0.4)  & 5.2 (0.6) & 2.2 (0.2)  & 7.4 (0.7)  \\ \\
%
%
  & ~~~~IV) $[4]_{X}[31]_{S}$: & &&&& && &  & &  & \\
 14 &  & 0 & 0 & 0 & 0 & 0 && 0 & 1.7 (0.2) & 1.7 (0.2) & 0 & 1.7 (0.2) \\
 15 &  & 0 & 1.0 (0.1) & 1.0 (0.1) & 1.1 (0.1)& 2.1 (0.2) && 1.9 (0.2) & 0.9 (0.1) & 2.8 (0.3) & 0 & 2.8 (0.3) \\
 16 &  & 1.0 (0.1) & 0.3 (0.0) &  1.3 (0.1) & 0.3 (0.1) & 1.6 (0.2) && 0.5 (0.1) & 0.3 (0.0) & 0.8 (0.1) & 0.6 (0.1)  & 1.4 (0.2) \\
 17 &  & 0 & 0.1 (0.0) & 0.1 (0.0) & 0 & 0.1 (0.0) && 0 & 0.1 (0.0) & 0.1 (0.0) & 0 & 0.1 (0.0) \\
   & ~Category~IV & 1.0 (0.1) & 1.4 (0.1) & 2.4 (0.3) & 1.4 (0.1) & 3.8 (0.4)   && 2.4 (0.3)  & 3.0 (0.3)  & 5.4 (0.6) & 0.6 (0.1)  & 6.0 (0.6)  \\ \\
   & All configurations & 10.0 (1.1) & 16.1 (1.7) & 26.1 (2.8) & 6.4 (0.7)& 32.5 (3.5) && 12.9 (1.4) & 13.5 (1.5) & 26.4 (2.9) & 5.6 (0.6)  & 32.0 (3.5) \\ \\
\hline
\hline
\end{tabular}
\end{table}
\end{appendix}

\end{document}